\documentclass[runningheads,a4paper,english]{llncs}[2022/01/12]
\usepackage[ngerman,main=english]{babel}
\usepackage[hyphens]{url}
\usepackage[T1]{fontenc}
\usepackage{graphicx}
\usepackage{xcolor}
\usepackage{dirtytalk}

\begin{document}

\title{A Copernican Revolution in Data}

\author{Claudio Gutierrez}
\vspace{-0.4cm}
\institute{Department of Computer Science, Universidad de Chile \& IMFD}
\maketitle

\vspace{-0.7cm}
\begin{abstract}
Half a century ago, Charles Bachman foresaw the significance and centrality of data in the digital world. In this short paper, we delve into the evolution of these ideas within the database community over the past decades. We believe that this historical analysis helps deepen our comprehension of the fundamental changes undergoing our discipline and provides insights into the future trajectory of our field.
\end{abstract}

\paragraph{\bf Data and Information Half a Century Ago.}
\label{sec:introduction}

In 1973, Charles Bachman earned the distinction of receiving the first Turing Award in the arena of data, acknowledging his groundbreaking contributions. In his speech, he said that a profound change in approaches to data management was needed, comparable to the momentous impact that Copernicus' heliocentric model had on astronomy \cite{bachman}. 
Bachman emphasized the importance of treating data as a central object of concern, while computer technology served merely as its servant. According to him, the traditional approach, which placed computers at the center, no longer suited modern times. This mindset required modification so that society might fully exploit the potential of digitizing business records, newspapers, books, and other forms of documentation. 
At that time, database management primarily focused on maintaining large amounts of structured data securely and efficiently \cite{codd}, whereas information retrieval involved searching and recovering textual content \cite{harman}. 

Inspired and guided by Bachman's insights, we explore below how the role of data has changed over the last few decades. Initially serving as a modest support to digital computers, data has now become a fundamental element in modern technology, society, and everyday life. This revolution in the role of data poses several challenges to the data discipline. In the conclusion, we highlight some of these challenges to encourage discussions within the data community.\footnote{
This paper was submitted to the Alberto Mendelzon Workshop on the Foundation of Databases, AMW, May 2023. It was well received by the reviewers but was considered out of scope. I thank Open Assistant Chat (https://open-assistant.io/chat) and ChatGPT (https://chat.openai.com/) for their assistance with English phrasing.
}


\paragraph{\bf 1970's: Data at the Center of Computing.}

According to Newell, Perlis, and Simon's assertion in 1967, computer science in the 1960s predominantly revolved around computers \cite{newell-perlis-simon}. 
Nevertheless, certain individuals observed that data held a prominent position in the field. In 1966, Peter Naur put forth the idea of substituting the term \say{Datalogy} (the study of the properties and uses of data) for \say{Computer Science} \cite{naur}. This suggestion highlighted the crucial function data plays alongside computation.

Bachman, who specialized in Information Systems, outlined a detailed plan for a paradigm shift from a computer-centric perspective to a database-driven approach, which he dubbed \say{a new basis for understanding [the area].} He maintained that emerging technologies allowed for a transition from sequential file technology to direct access storage devices, enabling users to perceive databases as interconnected networks of data and grant equal access to all its facets [key data fields]. According to Bachman, application programmers must adopt a \say{database-centric view} by navigating within a multidimensional data space.

To accomplish this transformation, Bachman believed that a novel science centered on seeking \say{minimal energy solutions for database access} was necessary. 
Comparing this pursuit to optimizing planetary movements (\say{Can you imagine restructuring our solar system to minimize travel time between planets?}),
he underscored the need for efficient database design. 

Bachman's ideas sparked advancements in information management, leading to the growth of Database Management Systems (DBMS) and solidifying the foundation of the database field. However, full implementation required approximately twenty years before achieving widespread recognition in 1988 (according to the Laguna Beach Report \cite{laguna-beach}). At that time, most database research revolved around improving DBMS performance, security features, and persistence, with attention primarily directed towards isolated systems. A year later in 1990 (Lagunita Report \cite{lagunita}), experts acknowledged that innovative applications such as scientific databases, design databases, and accessible global information could benefit immensely from incorporating database techniques. Overall, the evolution of data availability forced professionals to confront complex issues related to diverse data types across varied domains and locations.

\paragraph{\bf 1990s: Data Networked at World Scale: The Web.}

In the early 1990s, data became widely available on a global scale through the internet. One significant event during this era was the introduction of the World Wide Web, created in 1993 by Tim Berners-Lee. According to his vision, \say{The concept of the Web integrated many disparate information systems, forming an abstract imaginary space where the differences between them did not exist} \cite{web}.

Initially, the database community approached the Web with caution but eventually recognized its impact. By 1995, the Lagunita II report \cite{lagunita2} focused its research agenda on core DBMS concepts, \say{fundamental to current and developing information management needs}.
However acknowledged the \say{explosion of digitized information} over the previous five years. 
The report included discussions on the Web and the information explosion it brought, expecting that \say{the provision and use of such [Web] information would become a concern of each individual.} They foresaw that database technology would play a key role in this.

Shortly after, the Asilomar Report (\cite{asilomar}, 1998) made a clear statement: \say{The Web changes everything} and advanced the thesis that \say{the Web is one huge database.} The main concerns were universal accessibility, integration of different data types, and the need to reconsider the database research agenda in terms of content and method. They called for \say{a redirection of the research community away from incremental work and towards new areas}.

Similarly, the Lowell Report (\cite{lowell}, 2003) embraced the new reality, stating that \say{Database needs are changing, driven by the Internet and increasing amounts of scientific and sensor data.} They recognized the need for a next-generation infrastructure capable of addressing the integration of text, data, code, and streams, as well as the challenges posed by multimedia, uncertain data, personalization, and privacy. Additionally, they observed, \say{A new form of science is emerging. Each scientific discipline is generating huge data volumes.}

\paragraph{\bf 2008-today: From Big Data to a World of Data.}

At the start of the 21st century, the landscape for data research had significantly shifted. In 2006, Alexander Szalay and Jim Gray issued a warning to the scientific community about the challenges of an exponentially growing volume of data \cite{gray}. About the same time, the 2008 Claremont Report \cite{claremont} acknowledged the \say{excitement over big data} and recognized that its pervasiveness would fundamentally alter the database research field. Looking ahead, the authors concluded that \say{database research and the data-management industry area at a turning point}.

Five years later, in 2013, the Beckman Report \cite{beckman} identified big data as \say{a defining challenge of our time}. It  
described the driving forces behind this new paradigm: (1) the decreasing costs associated with generating diverse data, thanks to inexpensive storage, sensors, smart devices, social software, multiplayer games, and the emerging Internet of Things, which connects homes, cars, appliances, and other devices; (2) the reduced cost of processing vast amounts of data, thanks to advancements in multicore CPUs, solid-state storage, inexpensive cloud computing, and open-source software; and (3) the democratization of data, where not only database administrators and developers, but also decision-makers, domain scientists, application users, journalists, crowd workers, and everyday consumers have become intimately involved in generating, processing, and consuming data. As a result, \say{unprecedented volumes of data can be captured, stored, and processed, and the knowledge gleaned from such data can benefit everyone: businesses, governments, academic disciplines, engineering, communities, and individuals}.

Bachman's revolutionary vision in information systems foreshadowed the pivotal role data would play in various aspects of human affairs. In line with this perspective, the Seattle Report (\cite{seattle}, 2018) diagnosed the current state of affairs, stating that \say{data is at the center of everything today}, which has consequently led to the field's expansion and the emergence of new challenges.

\paragraph{\bf Revisiting Bachman's Thesis.}

Fifty years after Bachman's groundbreaking work on the significance of data in information systems, we have come to the realization that his insights not only apply to technology but also extend to human affairs. These advances have led to data transcending traditional disciplinary boundaries. Consequently, we can raise the question, as noted by Duguid, \say{Now that the banks have broken, should we still hope to have the river?} (see \cite{duguid}).

This evolving scenario presents numerous challenges and threats to our discipline.  To foster discussions within the AMW community, here are several concerns we wish to raise:

\begin{enumerate}

\item 
  Our field is becoming more interdisciplinary, making it difficult to define its scope. As a result, other fields consider data management their area of expertise without fully understanding it. This ambiguity results in an identity crisis that permeates research agendas, disciplinary programs, career paths, and the job market.

\item 
 With data now at the center of everything and the immense capacity to process it on a massive scale, highly profitable and impactful technologies, such as AI, have emerged. This raises a crucial question: In this context, what drives our research as DBMS did 50 years ago?

\item  Given our role as stewards of the digital world's content, we must address and
contribute to the destiny and ethical use of this \say{material} by society and organizations. This has given rise to various new areas of focus, including data governance, data privacy, data use policy, data sharing, and data ethics \cite{seattle}.

\item 
For scholars working on developing the discipline from Latin America, local concerns deserve attention. Although globalization may create a sense of homogeneity, specific regional matters, such as data sovereignty and threatened datasets from indigenous languages, require consideration today. 

\end{enumerate}

 

\bibliographystyle{plain}
  \bibliography{amw2023}
%
\end{document}